# Unconventional magneto-transport in ultrapure PdCoO$_2$ and PtCoO$_2$

Nabhanila Nandi[1], Thomas Scaffidi[2], Pallavi Kushwaha[1, †], Seunghyun Khim[1], Mark E. Barber[1], Veronika Sunko[1,3], Federico Mazzola[3], Philip D. C. King[3], Helge Rosner[1], Philip J. W. Moll[1], Markus König[1], Joel E. Moore[2], Sean Hartnoll[4] and Andrew P. Mackenzie[1,3]

[1]Max Planck Institute for Chemical Physics of Solids, Nöthnitzer Straße 40, 01187 Dresden, Germany

[2]Department of Physics, University of California, Berkeley, California 94720, USA

[3]Scottish Universities Physics Alliance, School of Physics and Astronomy, University of St. Andrews, St. Andrews, Fife KY16 9SS, United Kingdom

[4]Department of Physics, Stanford University, Stanford, California 94305-4060, USA

†Present address: CSIR-National Physical Laboratory, Dr. K.S. Krishnan Marg, New Delhi 110012, India

Correspondence: Andy.Mackenzie@cpfs.mpg.de

**Abstract**

We have studied magnetotransport in the single-band, quasi-two dimensional metals PdCoO$_2$ and PtCoO$_2$, which have extremely long mean free paths. We observe a strong temperature dependence of the Hall resistivity in small applied fields, linked to a large violation of Kohler's rule in the magnetoresistance. We discuss the extent to which these observations can be accounted for by standard transport theory, and describe other possible, unconventional contributions to magnetotransport in very high purity metals.

**Introduction**

Recent years have seen a rapid growth of research on unusual regimes of electronic transport (*1*). Of particular interest is one in which the mean free path deduced from standard measurements of electrical resistivity is very long. As is well known from the physics observed in high purity semiconductor two-dimensional electron gases and graphene, a host of novel physics can be seen in this high purity limit (*2–4*). In bulk systems, there has been a renaissance of research into ultrapure Bi (*5*), and the extremely low resistivity observed at low temperatures in compound metals such as $Cd_3As_2$, NbP and $WP_2$, is also generating considerable attention (*6–8*).

The majority of the bulk metals with extremely long mean free paths are either low carrier density semi-metals or materials with fairly complex multi-sheet Fermi surfaces and both electron- and hole-like carriers. It is highly desirable, therefore, to identify high carrier density metals with simple, single-sheet Fermi surfaces and long resistive mean free paths, to provide benchmark systems for understanding electrical transport. An additional benefit is quasi-two-dimensional conduction, to provide information at high carrier density to complement observations in the low density two dimensional systems.

The non-magnetic delafossite metals $PtCoO_2$ and $PdCoO_2$ satisfy all the above criteria. Conduction takes place in triangular lattice Pt and Pd layers, with a single conduction band crossing the Fermi level. The electrical conductivity of both materials is remarkably high. At room temperature $PtCoO_2$ has the longest mean free path of any mono- or divalent metal (longer even than those in elemental Cu or Ag) (*9*). Low temperature mean free paths of tens of microns have been observed in PdCoO$_2$ (*10*), and there is evidence that in this regime, there is strong phonon drag, in which the phonons gain net momentum when a current flows, meaning that normal electron-phonon scattering processes do not contribute to the observed resistivity (*10, 11*).

The extremely long relaxation times seen at low temperatures in PdCoO$_2$ have been exploited in studies of out of plane magnetoresistance (*12, 13*) and electrical flow in mesoscopic channels of restricted width (*14*), with the latter experiments uncovering evidence for hydrodynamic flow from analysis of zero field data. However, no comprehensive study of in-plane magneto transport has been reported. This is important both in its own right and because of recent theoretical work highlighting the possibility that hydrodynamic effects are in principle observable by comparing data from mesoscopic samples under applied magnetic fields with those seen in bulk samples (*15, 16*), experiments which require detailed knowledge of the bulk properties.

PdCoO$_2$ and PtCoO$_2$ are also attractive for study because of the remarkable simplicity of their basic electronic structure (*9*). The single, highly two-dimensional band with dominantly Pt or Pd character that crosses the Fermi level results in the Fermi surfaces shown in Fig. 1 A and B. Both have been extensively characterised by de Haas-van Alphen effect, angle-resolved photoemission measurements and electronic structure calculations (*10, 17–21*). The results of experiment and theory are in excellent agreement, as long as the correlations in the transition metal layers are taken into account so that the small deviations from perfect two-dimensionality due to interplane coherence are correctly reproduced (*19, 22*). Once this is done, both the calculations and direct photoemission experiments show that, in a two-dimensional approximation, the Fermi velocity is constant around the Fermi surface to within ± 5% (Fig. 1C). $PdCoO_2$ and $PtCoO_2$ are therefore ideal materials on which to test the predictions of transport theory.

**Results**

As described in the Methods section below, the extremely high conductivity of $PdCoO_2$ and $PtCoO_2$ means that devices with precisely defined geometries are required for accurate measurements; this was achieved by sculpting single crystals using focused ion beams. In total, 3 devices of this kind were made from $PdCoO_2$ and 3 from $PtCoO_2$ for this project; representative devices of each material are shown in Fig. 2 A and B respectively, with measured zero-field resistivity shown in Figs 2 C and D. Averaged across all measured devices, measured room temperature resistivity was 3.05 ± 0.07 μΩcm for $PdCoO_2$ and 1.82 ± 0.13 μΩcm for $PtCoO_2$, with intersample thickness variation the main source of experimental error. Given the precision with which the new devices were prepared, the above values should replace the previous best estimates [2.6 μΩcm (*10*) and 2.1 μΩcm (*19*)] for the room temperature resistivities of the two compounds. At low temperature the resistivity of the two samples shown in Fig. 2 is 8.1 nΩcm and 80 nΩcm, respectively at the low and high end of the ranges (8.1-29.3 nΩcm and 20-80 nΩcm) observed for the two compounds.

Throughout this paper we will discuss the analysis of transport data using expressions that, while commonly used, require precise definition. We adopt the following convention: 'Boltzmann' theory refers to solutions of the Boltzmann equation in the relaxation time approximation in which all scattering is momentum-relaxing, but the resulting mean free path can vary as a function of wave vector **k** around the Fermi surface. A 'Drude' expression is more restrictive, assuming the existence of a single, **k**-independent relaxation time. Since the Fermi velocity is so weakly **k**-dependent in $PdCoO_2$ and $PtCoO_2$ (Fig. 1C), a single relaxation time is equivalent to a single mean free path to a good approximation. The low temperature resistivity of the $PdCoO_2$ crystal (Fig. 2A) is particularly noteworthy because in a standard Drude analysis it corresponds to a mean free path of 20.3 μm and, therefore, to extremely weak momentum-relaxing scattering.

The Hall resistivity for the two samples is shown in Figs. 3 A and B for a range of temperatures between 2 K and 300 K. As emphasized by the derivative plots shown in Figs 3 C and D, there is a clear, temperature-dependent separation between low- and high- field behavior. The high-field derivative of approximately 0.027 μΩcm/T corresponds to a Hall coefficient ($R_H$) of $2.7 \times 10$ m$^3$/C, within a few per cent of the Drude expectation for Fermi surfaces of the volume established by the de Haas-van Alphen experiments. At low fields the value of $R_H$ drops, below a crossover field that is strongly temperature-dependent. In $PdCoO_2$ the crossover occurs at vanishingly small fields for $T < 30$ K, but its characteristic field grows rapidly as temperature is increasing, such that the high field regime is not reached within our range of applied magnetic fields unless the temperature is below approximately 150 K. Perhaps surprisingly for a single-band material with such a simple Fermi surface, the low field value of $R_H$ is as much as a factor of three smaller than the high field one. The measured magnetoresistance (MR) of $PdCoO_2$ and $PtCoO_2$ is shown in Figs. 4 A and B. At all temperatures, the overall scale of the MR is small, never exceeding 30% at 9 T.

**Discussion**

Independent of any specific framework of analysis, the data shown in Figs. 3 and 4 imply several things. Firstly, the transport must be controlled by more than one characteristic microscopic length scale. This can be deduced directly from the Hall effect data. The weak-field Hall coefficient can in principle differ from the high-field one because it is sensitive to details of scattering (*23*). However, if that scattering has only one characteristic length scale, i.e. the system is controlled by a single mean free path in a Drude picture, the length scale cancels from the expression for the weak-field Hall coefficient and the weak- and strong-field Hall coefficients must both equal 1/*ne* where *e* is the electronic charge and *n* the carrier density. The observation of the crossover in the Hall effect data therefore necessarily implies more than one length scale. Since the crossover field can be associated with a length set by the cyclotron radius $r_c = \hbar k_F/eB$ (where $\hbar$ is Planck's constant divided by $2\pi$ and $k_F$ is the average Fermi wave vector) which decreases with increasing field, the strong temperature-dependent increase of the crossover field implies that at least one of the microscopic length scales decreases rapidly with increasing temperature.

The existence of two length scales is also consistent with the observation of magnetoresistance, because within the Drude theory of a single-band material with only one microscopic length scale the MR vanishes. If there are two distinct scales, the MR will be non-zero and be related to their difference, while the resistivity in zero field will depend on some weighted average of the two. If the two scales have the same temperature dependence, MR data measured at different temperatures will collapse when plotted as $\Delta\rho(B)/\rho(0)$ against $[B/\rho(0)]^2$. This collapse, often referred to as Kohler's rule, is obeyed above approximately 150

K in PdCoO$_2$ and PtCoO$_2$, but is violated quite strongly below 150 K, as shown in Figs. 4 C and D. The Lorentz ratio has previously been measured in PdCoO$_2$ and shows analogous behaviour. Above 150 K it is temperature-independent, but develops a rather strong temperature dependence below 100 K (*11*).

The fact that the low temperature MR data fall below the Kohler-collapsed data implies that the separation of length scales that exists for $T > 150$ K is becoming smaller. This would be qualitatively expected to reduce the discrepancy between the weak- and high- field Hall coefficients, and this is clearly observed in the data: in Fig. 5A, we show the temperature dependence for $T < 150$ K of two factors, $\alpha(T) = \lim_{B \to 0} R_H(T)/R_H(2K)$ and $\beta(T) = \lim_{\frac{B}{\rho} \to 0} \frac{\Delta \rho/\rho}{(B/\rho)^2}(2K) / \frac{\Delta \rho/\rho}{(B/\rho)^2}(T)$. The similarity between the trends in $\alpha(T)$ and $\beta(T)$ is quantitative as well as qualitative, implying that a single temperature-dependent scale controls the weak field MR and Hall effect. The violation of Kohler's rule in this temperature range further implies that the scale that controls $\alpha(T)$ and $\beta(T)$ becomes significantly different from the scale that controls $\rho(0)$.

The data in Figs. 2-4, combined with Fig. 5A and the deductions made above, represent the main experimental results and model-free conclusions that we report in this paper. We close with a discussion of the extent to which they can be reconciled in a conventional picture of **k**-dependent scattering, and of ways in which less conventional scenarios might in principle contribute to magnetotransport in extremely pure metals.

Accounting for the observations in a conventional Boltzmann framework relies on the fact that the Fermi surfaces of Fig. 1 A and B are non-circular. The weak-field Hall coefficient $R_H \sim$ $\sigma_{xy}/\sigma_{xx}^2$ where $\sigma_{xy}$ and $\sigma_{xx}$ are the Hall and magnetoresistive conductivities respectively. In the relaxation time approximation with only standard momentum-relaxing scattering, the length scales referred to in the above discussions are expressed as mean free paths $\ell$. If there is only one value of $\ell$, $R_H$ is independent of scattering, because $\sigma_{xy} \sim \ell^2$ and $\sigma_{xx} \sim \ell$. However, the elegant geometrical construction of Ref. (23) for interpreting the weak-field Hall effect in two-dimensional metals highlights the fact that for Fermi surfaces around which the curvature varies, the high curvature regions dominate $\sigma_{xy}$ while an average around the whole Fermi surface determines $\sigma_{xx}$. If the mean free path $\ell_1$ on the high curvature regions is smaller than that ($\ell_2$) on the low curvature regions, and the curvature changes strongly around the Fermi surface, $R_H$ is suppressed from its high field value by a factor $\gamma(\ell_1/\ell_2)^2$ where constant $\gamma$ is related to the precise Fermi surface geometry. Modelling a hexagon with corners of varying curvature as described below in Methods, it is straightforward to show that the required mean free path anisotropy in the delafossites is approximately a factor of two (Fig. 5B).

The observation of weak field MR in a material with a single two-dimensional Fermi surface requires, within Boltzmann theory, that $\ell$ varies around that Fermi surface (24). Some MR is expected, therefore, if $\ell_1 \neq \ell_2$, with its value being determined by details of the change from $\ell_1$ to $\ell_2$ around the Fermi surface. If the ratio $\ell_1/\ell_2$ becomes smaller at low temperatures, the weak field Hall coefficient rises and the magnitude of MR is qualitatively expected to fall. The close correlation evident in Fig. 5A would, however, be the result of some fine-tuning.

Although the above discussion shows that it is possible to construct a model within conventional Boltzmann theory that can capture the main features of our observations, a microscopic justification would be required for at least two of the necessary ingredients of the model. Firstly, it is not clear why there should be a factor of two change in $\ell$ around two-dimensional Fermi surfaces around which the changes of $v_F$ are much smaller, and are oriented differently between $PdCoO_2$ and $PtCoO_2$ (see Fig. 1C). Secondly, it is far from obvious that the ratio $\ell_1/\ell_2$ should be so strongly temperature dependent below 150K.

Although the conventional analysis discussed above is capable of capturing a number of the qualitative features of our observations, assumptions and parameter tuning are required to make it work. More generally, the fact that both the Fermi surface shape and $v_F$ variation differ in detail between $PdCoO_2$ and $PtCoO_2$ is in tension with an explanation of their similar magneto-transport properties that essentially makes use of detail. For these reasons, it is useful to consider unconventional possibilities.

At low $T$, both the (1) longitudinal and (2) Hall resistivities point towards a long momentum-relaxing mean free path, and therefore a potentially hydrodynamic regime of transport: (1) the longitudinal resistivity is smaller than expected from Kohler's rule, and (2) the Hall effect takes its universal value, $R_H = 1/ne$, even at low fields, in agreement with the hydrodynamic prediction, which is independent of Fermi surface shape (*25–28*). This motivates us to discuss how certain features of electron hydrodynamics can in principle play a role in the present system.

Naively, one might think that momentum-conserving scattering should be irrelevant for bulk transport properties, and indeed the study of hydrodynamic effects in electronic transport has been so far limited to shear viscous effects in mesoscopic samples (*14, 29, 30, 3, 31–34*). However, there are at least two ways by which momentum-conserving scattering can impact

X

transport in a bulk crystal. Firstly, there are other possible hydrodynamic effects which are seldom discussed because they are ruled out in Galilean invariant systems but should be present in any solid system with a non-circular (and therefore non-Galilean-invariant) Fermi surface such as those shown in Fig. 1. In the absence of Galilean invariance, conserving quasiparticle momentum is not the same as conserving overall current, so momentum-conserving processes can in principle have an effect on bulk properties. Specifically, in a general homogeneous hydrodynamic fluid current **j** = $n$**v** + $\sigma_0$(**E** + **v** × **B**). Here, the first term is the carrier density times the fluid velocity, while the 'incoherent' transport coefficient $\sigma_0$, arising directly from the electromagnetic fields **E** and **B**, vanishes with Galilean invariance. The effect of $\sigma_0$ on magnetotransport is discussed in (*25–28*). The scale of the hydrodynamic contribution for $PdCoO_2$ and $PtCoO_2$ has not been calculated and may be small, but it should in principle be considered. Secondly, even nominally very pure materials can have long-wavelength disorder, which can in some circumstances act like a flow restriction and give rise to viscous effects becoming visible even in bulk crystals, in the same way as a viscous fluid flows through a porous medium (*35, 36*). Interestingly, in this setting or in the presence of a significant $\sigma_0$, the Hall component of the viscosity tensor would be expected to lead to a softening of the Hall slope at low fields, as observed in Fig. 3a (*15, 16*).

We believe that our observations motivate further theoretical work in this area, to establish the extent to which hydrodynamic contributions influence magnetotransport in extremely pure bulk metals. Whatever their explanation, the main findings of this paper are the experimental results reported in Figs. 2 to 4, in which the components of the magnetoresistivity tensor of bulk $PdCoO_2$ and $PtCoO_2$ are determined over wide ranges of temperature and magnetic field, and Fig. 5A, which shows that the temperature dependence of the weak field Hall coefficient tracks the violation of Kohler scaling in the magnetoresistance.

**Conclusion**

In conclusion, we have measured the bulk in-plane Hall effect and MR of the ultrapure metals $PdCoO_2$ and $PtCoO_2$, by performing simultaneous multi-channel low-noise transport measurements on microstructured single crystals. Model-free examination of the data reveals the existence of two microscopic length scales, each with a strong temperature dependence. More detailed analysis in the standard relaxation-time approximation shows that aspects of the data can be reconciled to some degree with conventional theory, but more exotic contributions can also exist in principle, and we hope that our experimental findings motivate further theoretical work on these fascinating two-dimensional metals.

**Methods**

**Device preparation and measurement**

Single crystals of $PdCoO_2$ and $PtCoO_2$ were grown in sealed quartz tubes using methods discussed in (*19*, *37*). The focused ion beam sculpting of these crystals to well-defined defined device geometries was performed in a dual beam liquid gallium FIB (FEI Helios) using adaptations of methods described in detail in (*14*, *38*). Electronic band structure calculations were performed using methods described in (*19*, *22*), with the resulting Fermi velocities averaged across the $k_z$ direction of the Brillouin zone to produce the two-dimensional projections summarized in Fig. 1C.

Experimentally, nearly two-dimensional metals with extremely high electrical conductivity are in a fairly unusual regime, and measuring the intrinsic bulk transport properties requires special care. Injection of current through top contacts can result in an inhomogeneous depth distribution of current in the measurement channel; we avoid this by patterning in the long meanders between the current injection point and the channel, confirming with simulations and multi-contact measurements that homogeneous in-channel currents have been achieved. To probe the ohmic rather than the ballistic regime both the width of the sample and the spacing between voltage contacts should be as large as possible. To ensure that the magnetotransport data are a good approximation to the bulk limit, even when the mean free path is as large as 20 $\mu$m, large channel widths (155 $\mu$m and 190 $\mu$m for PdCoO$_2$ and PtCoO$_2$ respectively) and longitudinal voltage contact separation (204 $\mu$m and 244 $\mu$m or PdCoO$_2$ and PtCoO$_2$ respectively) were used. However, for as-grown crystals of PdCoO$_2$ with a typical thickness 10-20 $\mu$m, wide crystals have very low resistance at low temperatures, so good voltage resolution is required. Another issue is the relative scale of resistance and Hall resistance. In most metals in standard configurations for transport measurements, the resistance is much larger than the Hall resistance for magnetic fields $B$ in the range -10 T < $B$ < 10 T, but in these delafossites the situation is reversed, and even tiny thickness variations can lead to a pronounced odd-in-field contribution to the magnetoresistance. Surprisingly, an apparent odd contribution to magnetoresistance does not violate Onsager's relations in exotic cases for which transport is non-local (*39*), so differentiating between this possibility and effects caused by non-ideal crystal shapes requires precise sample preparation and geometrical characterization. For that reason we performed the measurements described here on crystals carefully selected for uniform thickness, and sculpted to well-defined geometries using focused ion beam milling. Measuring thickness across large samples is not always easy, so all measurements were made in multi-contact configurations to check for consistency. To avoid this becoming prohibitively time-consuming we designed a bespoke probe and readout system featuring a Synktek 10 channel lock-in amplifier, and studied all relevant configurations simultaneously with a voltage noise level of 1.5 nVHz$^{-1/2}$ using standard a.c. methods at a measurement frequency of 73.3 Hz for PdCoO$_2$ and 177.7 for PtCoO$_2$ and current of 1 mA. Fine temperature control and all readout was achieved with this home-built system; field and coarse temperature control were obtained by mounting the measurement probe in a 9 T Quantum Design Physical Property Measurement System.

**Modelling the low-field Hall slope**

The ratio of the low field Hall slope to the high field Hall slope is given by [ref]

$$r = \frac{\Gamma A_\ell}{\pi \ell_{av}^2} \quad (1)$$

where $\Gamma = \frac{4\pi A}{S^2}$, $A$ is the area enclosed by the Fermi surface, $S$ is the perimeter of the Fermi surface, $A_\ell$ is the area enclosed by the "$\ell$-surface" (see [Ong]), and $\ell_{av}$ is the average mean free path:

$$\ell_{av} = \frac{1}{S} \int ds \ell(s) \quad (2)$$

where *s* is a parametrization of the Fermi surface and $\ell(s)$ is the mean free path at point *s*.

We consider a rounded hexagon for the Fermi surface, with side *c* and radius of curvature *R* (Fig. 6). This leads to

$$S = 6\left(c - \frac{2R}{\sqrt{3}} + \frac{R\pi}{3}\right)$$

(3)

$$A = \frac{3\sqrt{3}c^2}{2} + \pi R^2 - \frac{6}{\sqrt{3}}R^2$$

We use a simple model for a momentum-dependent mean free path:

$\ell(\mathbf{k}) = \ell_0$ on flat edges

(4)

$\ell(\mathbf{k}) = \delta\ell_0$ on rounded corners

This leads to

$$\ell_{av} = \frac{\left(c - \frac{2R}{\sqrt{3}}\right)\ell_0 + \left(\frac{R\pi}{3}\right)\delta\ell_0}{\left(c - \frac{2R}{\sqrt{3}} + R\pi/3\right)}$$

(5)

and

$$A_\ell = \pi(\delta\ell_0)^2$$

(6)

Combining everything, we find the ratio of the low field Hall slope to the high field Hall slope to be

$$r = \Gamma\delta^2 \left(\frac{\left(1-\frac{2\eta}{\sqrt{3}}\right)+\left(\frac{\eta\pi}{3}\right)\delta}{\left(1-\frac{2\eta}{\sqrt{3}}\right)+\left(\frac{\eta\pi}{3}\right)}\right)^{-2} \qquad (7)$$

where $\eta = R/c$.

**Acknowledgements**

We gratefully acknowledge the technical assistance of S. Seifert, and interesting discussions with P. Surowka, P. Witkowski and R. Moessner.

**Funding**

We acknowledge the support of the Max Planck Society, support from the European Research Council (through the QUESTDO project), and the Engineering and Physical Sciences Research Council, UK (grant no. EP/I031014/1). T.S. acknowledges support from the Emergent Phenomena in Quantum Systems initiative of the Gordon and Betty Moore Foundation and V.S. thanks EPSRC for PhD studentship support through grant number EP/L015110/1.

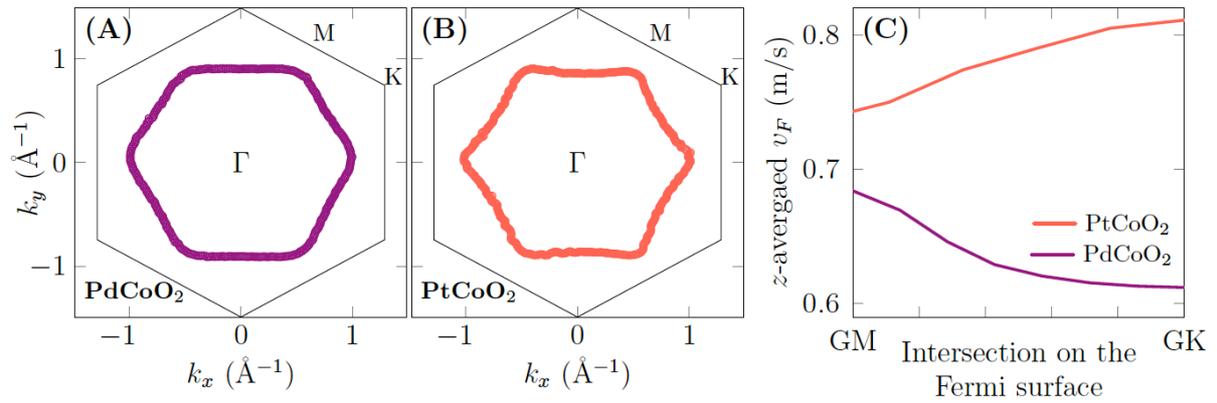

**Fig. 1** The Fermi surfaces of PtCoO$_2$ (**A**) and PdCoO$_2$ (**B**) as determined by angle-resolved photoemission. Both are highly two-dimensional and nearly hexagonal, with slight differences in curvature between the two compounds. **C** The Fermi velocities $v_F$ for the two compounds, obtained from electronic structure calculations optimized to match the experimentally determined band parameters, projected into a two-dimensional Brillouin zone. In both cases $v_F$ varies around only by ± 5% of its mean value, with the maximum along Γ- K in PtCoO$_2$ and Γ- M in PdCoO$_2$.

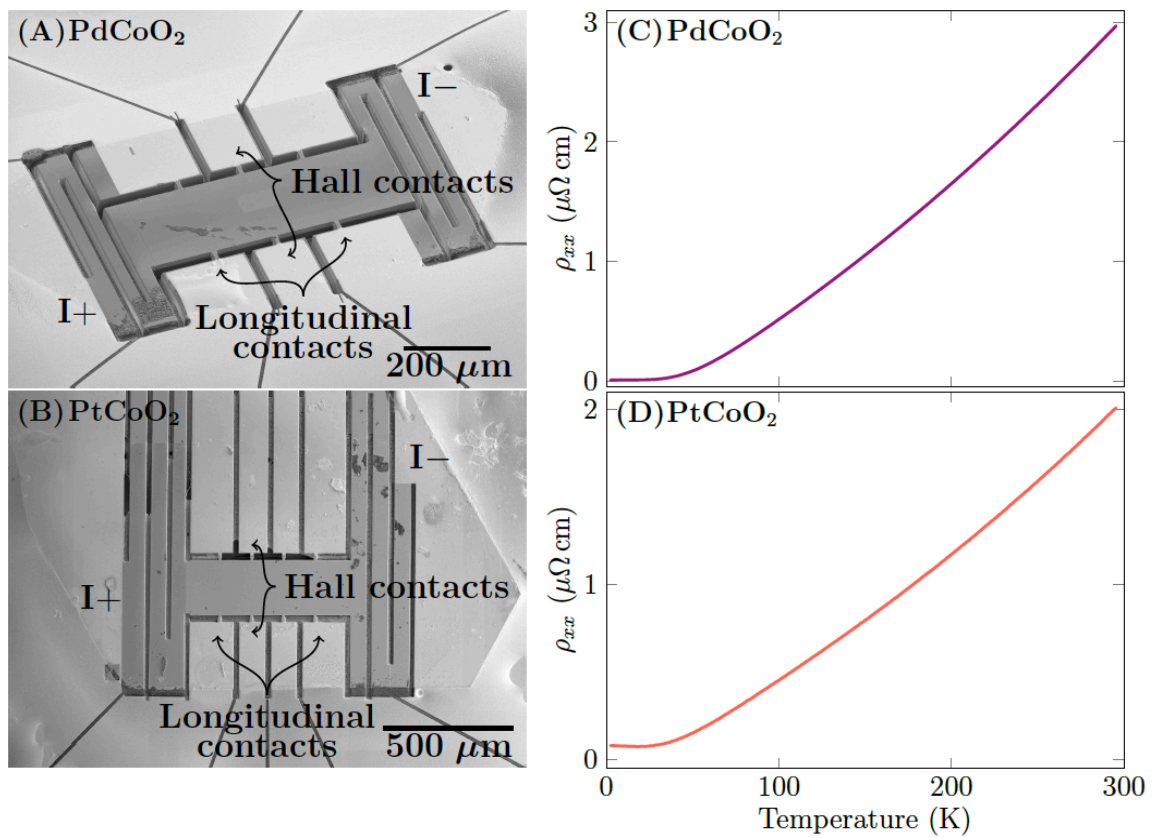

**Fig. 2** The resistivity of the PdCoO$_2$ (**A**) and PtCoO$_2$ (**B**) devices shown in panels **C** and **D** respectively. Each device was sculpted from a single crystal using focused ion beam milling. Electrical contact is made to the crystal via a layer of sputtered gold applied to the top surface. To ensure current homogeneity through the thickness of the measurement channel the current passes along a meander track before entering that channel. The longitudinal and Hall contact pairs for which $\rho_{xx}(H)$ and $\rho_{xy}(H)$ data are shown in this paper are marked, but the multi-contact geometry allowed for checks that these data were consistent across each sample.

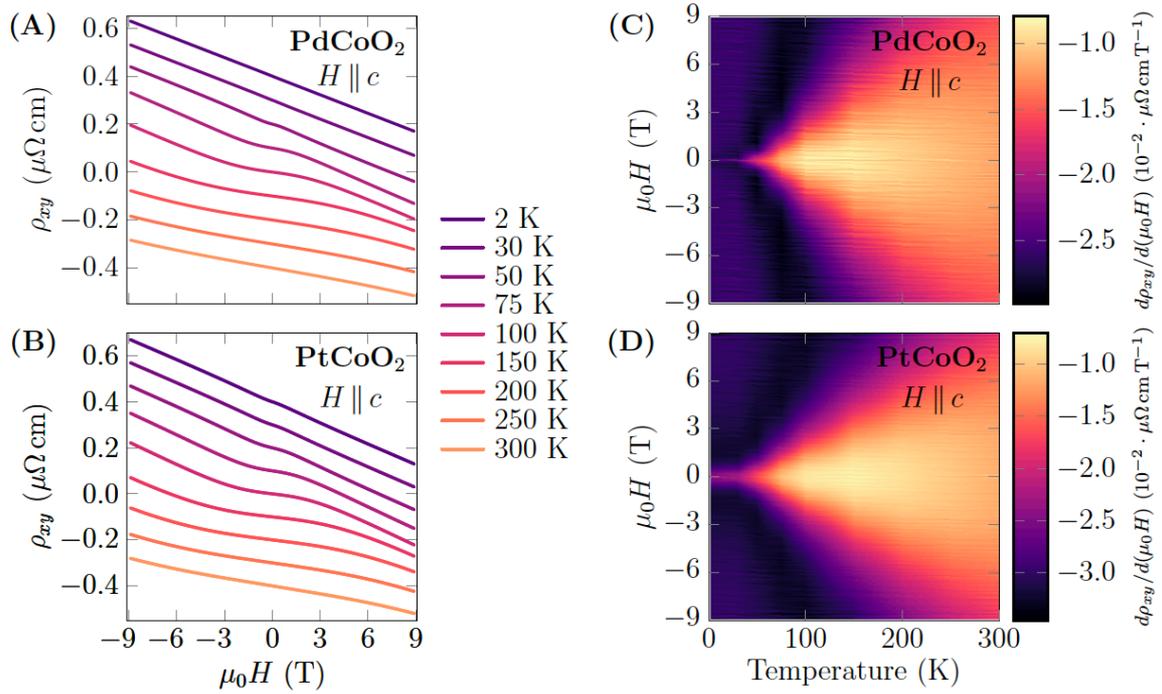

**Fig. 3** The Hall resistivity $\rho_{xy}$ for PdCoO$_2$ (**A**) and PtCoO$_2$ (**B**) at nine temperatures between 2K and 300 K, offset for clarity. A pronounced decrease in gradient is seen at low fields, with a characteristic field width that grows rapidly with temperature. This is illustrated in panels **C** and **D**, showing colour scale plots of d$\rho_{xy}$/dH as a function of field and temperature.

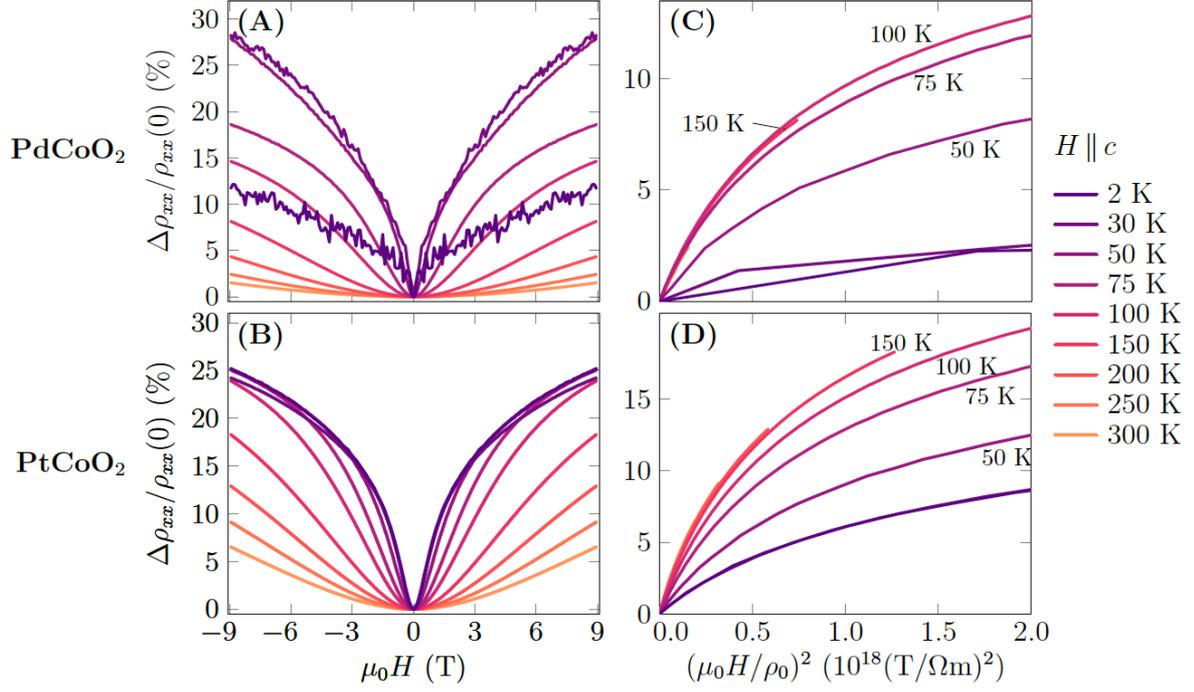

**Fig. 4** The magnetoresistance of PdCoO$_2$ (**A**) and PtCoO$_2$ (**B**) at nine temperatures between 2K and 300 K, expressed in the conventional manner as $\Delta\rho/\rho$. Voltage noise levels in all cases were 1.5nV/√Hz in the 10 channel lock-in setup that we were using; the higher noise levels on the 30 K and 2 K data reflect the extremely low resistance of the PdCoO$_2$ sample when its resistivity falls to 10 nΩcm and below. In panels **C** and **D** we show the strong violation of Kohler's rule below 100 K and 150 K in PdCoO$_2$ and PtCoO$_2$ respectively.

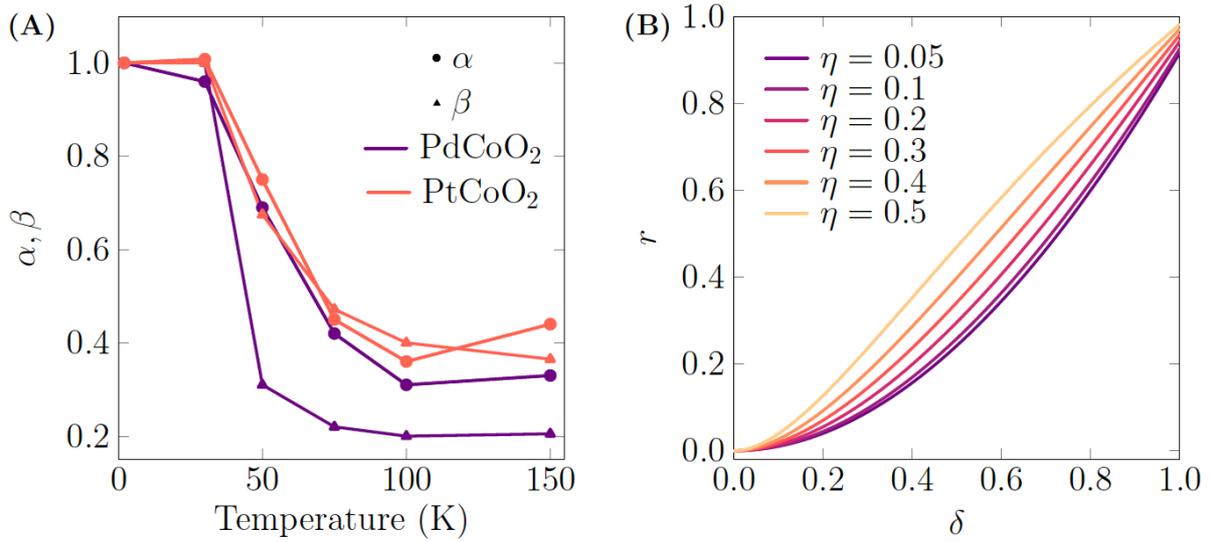

**Fig. 5 A** The factors α (circles) and β (triangles) as defined in the main text for PtCoO$_2$ (orange) and PdCoO$_2$ (purple) respectively. Data for β for PdCoO$_2$ are included for completeness, but are less reliable because the of extremely low resistivity at low temperatures leads to larger relative error in its estimation. In panel **B** we show the results of model calculations for the ratio *r* of the weak- to strong-field Hall coefficients predicted for a Fermi surface approximated to a hexagon with rounded corners. The experimentally observed value of *r* = 0.3 can phenomenologically be accounted for by a factor δ suppressing the mean free path on the Fermi surface corners relative to that on the faces. Depending on the value chosen for the curvature factor η, δ varies between 0.3 and 0.6.

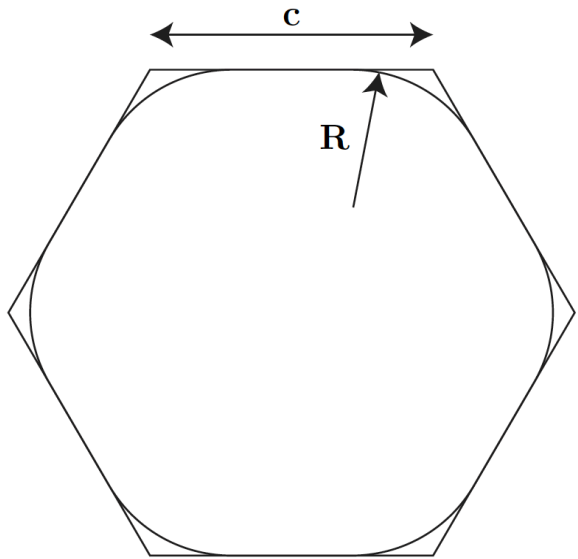

**Fig. 6** Sketch showing how parameters *c* and *R* of the model described in the Methods section are defined.